\documentclass{supp}
\usepackage{graphicx}

\title{Supplementary Information for
``A \boldmath{$15.65\,M_{\odot}$}
black hole in an eclipsing binary in the nearby spiral galaxy Messier 33''}

\author{Jerome A. Orosz$^{1}$, 
Jeffrey E. McClintock$^2$, 
Ramesh Narayan$^2$,
Charles D. Bailyn$^3$,
Joel D. Hartman$^2$,
Lucas Macri$^4$,
Jiefeng Liu$^2$, 
Wolfgang Pietsch$^5$, 
Ronald A. Remillard$^6$,
Avi Shporer$^7$ \&
Tsevi Mazeh$^7$
}

\begin{document}

\def\lesssim{\lower2pt\hbox{$\buildrel {\scriptstyle <}
   \over {\scriptstyle\sim}$}}

\maketitle

\begin{affiliations}
\item Department of Astronomy, San Diego State University,
5500 Campanile Drive, San Diego, CA 92182-1221, USA.
\item Harvard-Smithsonian Center for Astrophysics, 60 Garden Street,
Cambridge, MA 02138, USA.
\item Department of Astronomy, Yale University,
PO Box 208101, New Haven, CT 06520-8101, USA.
\item National Optical Astronomy Observatory, 950 North
Cherry Avenue, Tuscon, AZ 85719, USA.
\item Max-Planck-Institut f\"ur extraterrestrische Physik,
Giessenbachstra\ss e,  D-85741 Garching, Germany.  
\item MIT Kavli Institute for Astrophysics and Space Research, 
77 Massachusetts Avenue, 37-287, 
Cambridge, MA 02139, USA.  
\item Wise Observatory, Tel Aviv University,  Tel Aviv 69978, Israel.  
\end{affiliations}

\parindent 0pt
This Supplementary Information provides details about the spectral
extraction (crowding issues and the removal of nebular lines),
a discussion about the distance to M33, a model for the O-star
wind and the measurement of the true photospheric X-ray eclipse width
$\Theta$, 
and details about ellipsoidal modelling.  It also contains
five related figures, one related table, and additional references.

\parindent 0.39in

\section{Supplementary methods}

\subsection{Crowding issues:}

M33 X-7 lies in a moderately crowded OB association known as HS 13
(ref.\ 23).
Fortunately, HS 13 has been  observed
using the {\em Hubble Space Telescope} and the WFPC2.
In ref.\ 5 it is 
shown that apart from a close pair
(0.2 arcsecond separation) of stars 0.9 arcseconds to the southwest,
the optical counterpart of M33 X-7 is relatively isolated.
For the spectroscopic observations from the Gemini-North Telescope, the
0.5 arcsecond wide slit was placed at a position angle of 215.6 degrees,
which is the angle defined by M33 X-7 and the close pair of neighbour stars,
which are the only potential source of contamination.

\begin{figure}
\centerline{\includegraphics[scale=0.72,angle=-90]{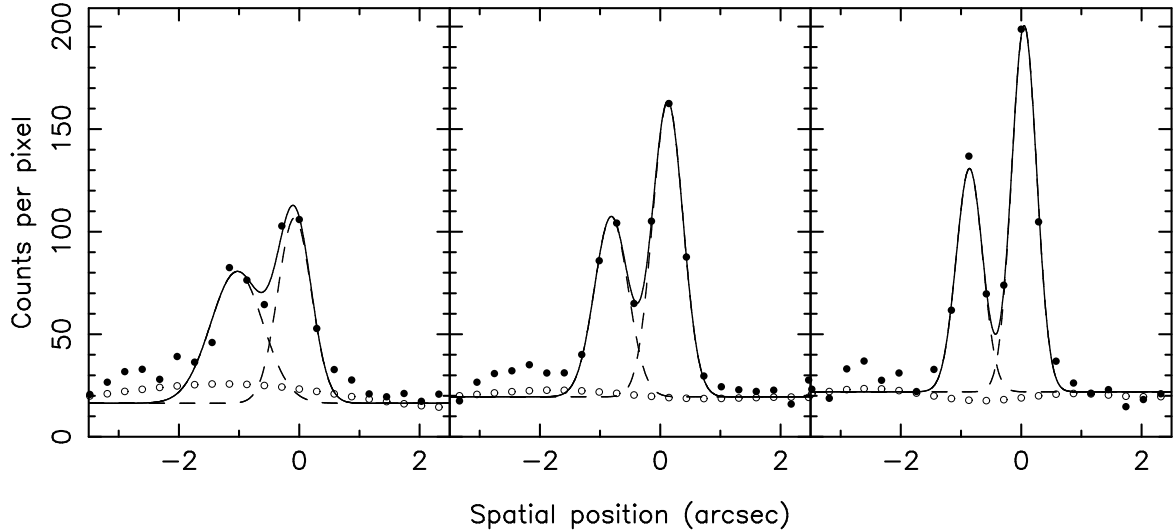}}
\caption{{\bf $\vert$ Spatial profiles of M33 X-7 and a close pair of
neighbour stars.} 
Shown are the profiles (filled circles)
of the spectrum in the spatial
direction near H$\beta$
of M33 X-7 (taller profile) and the neighbour
stars for three observations with different seeing conditions:
$\approx 0.8$ arcseconds full width half maximum (FWHM), left panel,
$\approx 0.6$ arcseconds FWHM, middle panel, and
$\approx 0.4$ arcseconds FWHM, right panel.  Most of the observations
were similar to the one shown in the middle panel.  The solid line
is a double Gaussian fit to the two profiles.
The dashed lines show the individual
Gaussians.  In all cases, the overlap region between the
two Gaussians is relatively small.  The area in the overlap region is,
from left to right, 18.6\%,
6.3\%, and 2.4\%  of the area under the profile of M33 X-7.
The open circles show the profile of the residual image computed by
SPECRES (e.g.\ the reconstructed spectrum minus the observed spectrum).
\label{plotcuts}}
\end{figure}

The detector in the GMOS instrument consists of
three $2048\times 4068$ EEV CCD chips in a row with $\approx 0.5$mm gaps
in between. The dispersion axis runs along the longer dimension
of the mosaic.  Supplementary 
Figure \ref{plotcuts}
shows spatial profiles
of M33 X-7 and the neighbour stars near H$\beta$ for three observations
that represent relatively ``poor'' seeing (about 0.8 arcseconds),
``average seeing'' (about 0.6 arcseconds), and ``good'' seeing
(about 0.4 arcseconds).    The majority of the spectra were similar
to the ``average'' case.  Gaussian fits to the spatial profiles
and simple numerical integration show that the area in the overlap
region of the two profiles is typically 5-10\% and 
at most 20\% of the area under
the M33 X-7 profile.  For the spectral extraction with the GMOS
pipeline, the lower extraction aperture was placed slightly to the
right of the ``dip'' between the two profiles, which was sufficient
to exclude light from the neighbour stars.

\subsection{Removing nebular lines:}

As noted above, M33 X-7 lies in an OB association, and light from
the surrounding HII region is evident in both the direct images
and in the two-dimensional spectra. 
Unfortunately, the background nebular light varies over small spatial
scales and  is not removed by the regular GMOS spectral extraction routines,
which assume that
the background light is uniform over the entire spatial
extent of the two-dimensional spectra.  Also, the presence of
the close pair of neighbour stars makes it harder to use higher order
polynomials when fitting the background region.

The SPECRES package in IRAF provides tools to extract point source
spectra with complex backgrounds$^{22}$.
The spatial profile of a point
source is distinguished from the quite different profile of an extended
background source using a Gaussian smoothing kernel along with a
Richardson-Lucy type of iterative restoration algorithm that performs a
maximum likelihood estimation.  For optimal results,
one must
precisely specify the positions of the point sources,
the variation of the spatial position of the point sources with
wavelength, and the
spatial ``Slit Spread Function'' (SSF) with wavelength.  
A list of the {\em relative} positions of the point sources
along the slit was made from the Gemini direct images and also
the archival {\em HST} images.  Given the list of relative positions,
the positions of the sources on a given observation can be found once
the position of a reference source is measured.
The variation of the position of the
spectra with wavelength (the ``slope'')
is easily measured using the extraction
tools in the GMOS package.
Unfortunately, the derivation of the
SSF is somewhat more involved since it depends on the properties of
the instrument and on the observing conditions.  The ``specpsf'' task
in SPECRES was employed
to make SSFs from the direct images that
were taken immediately following each spectroscopic observation
and by using analytic functions (Gaussians and Lorentzians) with 
widths that varied with wavelength.

Once the point source positions, the slope, 
and SSFs are determined, one must decide on a width for
the smoothing kernel that is used to distinguish between point sources
and extended background sources.  There is a trade-off between the
width of the kernel and the signal-to-noise (SN) in the extracted spectra.
A broad smoothing kernel will result in spectra with higher SN, but the
background subtraction will be poor if the background varies on small
spatial scales.  On the other hand, a narrow smoothing kernel will
result in spectra with good background subtraction, but at the expense
of lower SN in the extracted spectrum.

Since the observing conditions change from observation to observation,
the optimal set of parameters for one spectrum may not be appropriate
for another spectrum.  To optimise the spectral extraction for each
observation, we wrote scripts that performed
the spectral extraction with a wide range of
variations of the position of the
reference source, the spectral slope, the variation of the width of the SSF
with wavelength, and the size of the smoothing kernel.  For each observation,
372 extractions were performed, and the optimal extraction
was determined by cross correlation with a synthetic spectrum as the
template. 

\begin{figure}
\centerline{\includegraphics[scale=0.68,angle=-90]{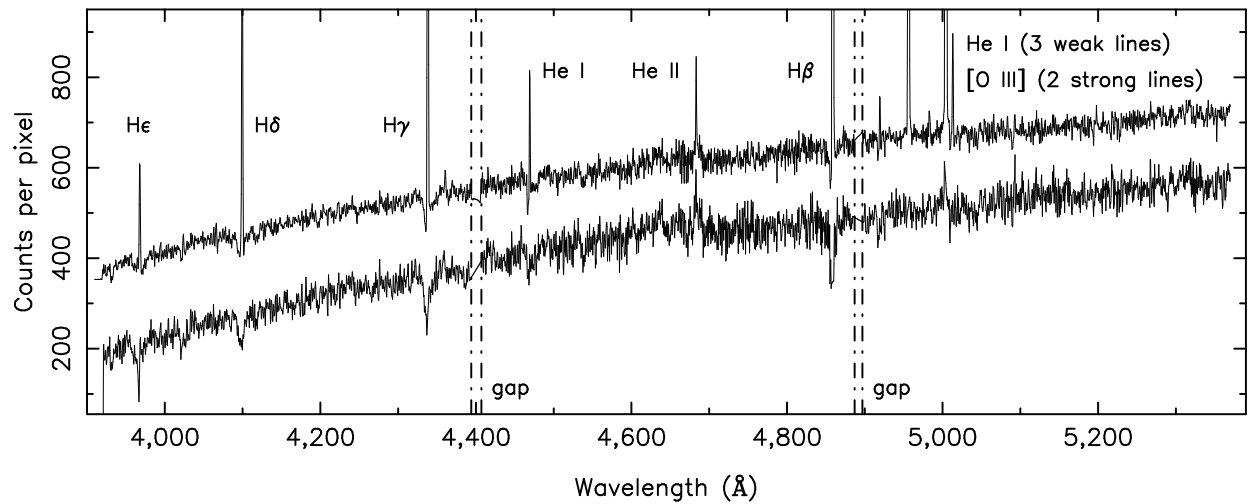}}
\caption{{\bf $\vert$
Comparison of GMOS pipeline and SPECRES extractions.}
({\it top}) Typical one-dimensional spectrum as extracted
using the GMOS pipeline (offset upwards by 200 units for clarity) with
nebular lines (e.g., H I, He I, and [O III]) present.   ({\it bottom})
Spectrum produced using SPECRES.  With the exception of 
H$\beta$ and the
stronger [O III] line, the nebular lines have been cleanly removed;
note, however, the increased noise level.  The regions of the spectrum
falling in the detector gaps are indicated by the vertical lines.
\label{specfig2}}
\end{figure}

With the exception of the stronger [O III] line and H$\beta$,
the nebular lines were cleanly removed in all
of the spectra.  Supplementary Figure \ref{specfig2}
shows
a spectrum extracted with the GMOS pipeline software and the spectrum
extracted with SPECRES.  
The nebular lines are mostly gone in the latter
spectrum, although note the higher noise level.

Since the H$\beta$ emission line was never cleanly removed, we
excluded this feature when modelling the spectrum and computing
the radial velocities.  As a check on the results, we note that the
radial velocity curve derived from the SPECRES extracted spectra
(using several Balmer lines and He I lines in the cross correlation
region
4000-4375 and 4450-4565~\AA) 
is virtually the same as the radial velocity curve derived
from the GMOS pipeline extracted spectra (using the two strong
He II lines noted in the main text).  
For example, for the former, we find $K_2 =112.0\pm 
7.6$ km s$^{-1}$ and for the 
latter we find $K_2 = 108.9 \pm 6.4$, 
which are in close
agreement with our adopted value in Table 2, main text.

\subsection{The distance to M33:}

  We estimate the distance to M33 using four indicators. Because of
the differential nature of distance measurements, we express our
intermediate results as relative distance moduli with respect to the
Large Magellanic Cloud (LMC), and then place the final result on an
absolute scale. 


{\em Tip of the Red Giant Branch}:  
Tip of the red giant branch
(TRGB)
magnitudes for ten fields in M33 are given in ref.\
26, 
and a global TRGB value for the LMC is derived in ref.\ 27. 
Both measurements were carried out
using the same calibration of the technique, which includes a small
($-0.03$~mag) metalicity correction. We find $\Delta\mu ({\rm
M33}-{\rm LMC})=6.22\pm 0.10$~mag.

{\em Cepheids}: 
$V$ and $I$ photometry of 61 Cepheids in
M33 with periods from 15 to 80 days is presented in ref.\ 28, 
which were fitted with LMC 
period-luminosity
(P-L)
relations from the OGLE project\cite{uda99}. 
We apply an
updated correction for the metalicity dependence of the Cepheid P-L
relation, based on the results 
in refs.\ 30 and 31, 
that amounts to $+0.06$~mag. Thus, we obtain $\Delta\mu ({\rm
M33}-{\rm LMC})=6.18\pm 0.03$~mag.

{\em RR Lyrae}: 
Sixty four RR Lyrae
variables in M33 discussed in ref.\ 32 were used to derive an average
extinction-corrected $V$ magnitude for the sample of
$\langle V\rangle_0=25.34\pm0.07$~mag. 
New
photometry and spectroscopy for over 100 objects in the bar of the LMC 
is presented in ref.\ 33, 
from which an average value of
$\langle V\rangle _0=19.07\pm0.06$~mag
was derived. While these values are already
corrected for the metalicity dependence of the RR Lyrae, the
calibrations used by the authors are slightly different 
(e.g.\ ref.\ 34 vs.\ ref.\ 35).
This results in an additional
$-0.02$~mag correction to the 
results in ref.\ 32,
bringing the
relative distance modulus to $\Delta\mu ({\rm M33}-{\rm
LMC})=6.25\pm 0.09$~mag.

{\em Red Clump}: 
An extinction-corrected $I$-band magnitude of the red clump in the
LMC of $I_0=18.12\pm 0.06$~mag
is given in ref.\ 33. 
An
equivalent value for M33 of $I_0=24.43\pm 0.04$~mag was determined in ref.\
26.
Both references use
two sets of calibrations for the red clump magnitude as a function of
metalicity\cite{uda00,pop00}
which
results in a small ($+0.03$~mag) metalicity correction   and a relative
distance modulus of $\Delta\mu ({\rm M33}-{\rm
LMC})=6.34\pm 0.07$~mag.

The average distance modulus of these four techniques, weighted by
their uncertainties, is 
$\Delta\mu ({\rm
M33}-{\rm LMC})=6.21\pm 0.03$~mag.

We calculate the true distance modulus of the LMC based on the
weighted average of three recent and
independent estimates. First, the
study of several detached eclipsing binary systems (DEBs)\cite{fit03}
yields an LMC distance modulus of
$18.42\pm 0.06$~mag. Second,
parallaxes of Galactic Cepheids obtained with the {\em Hubble Space
Telescope}
yield an absolute calibration
of the Cepheid P-L relation that, when applied to the LMC, gives a
distance modulus of $18.40\pm 0.05$~mag (after a correction for
metalicity dependence). Lastly, the relative Cepheid distance
modulus between NGC$\,$4258 and the LMC\cite{mac06}, 
coupled
with the maser distance to the former\cite{her99}, 
gives
$18.41\pm0.16$~mag. The average of these values (weighted by their
uncertainties) is 
$\mu{\rm (LMC)}=18.41\pm0.04$~mag.

In conclusion, the combination of $\Delta\mu ({\rm M33}-{\rm LMC})$
and $\mu{\rm (LMC)}$ yields $\mu{\rm (M 33)}=24.62\pm 0.05$~mag, which
corresponds to a distance of $d=840\pm 20$~kpc.

We note that 
our result is in disagreement with the recent distance determination
to M33 in ref.\ 40,
who obtain $\mu{\rm
(M33)}=24.92\pm 0.12$~mag, or $d=960\pm 50$~kpc based on a DEB.
This is a puzzling result, because it implies
a relative distance modulus between the LMC and M33 based on DEBs of
$6.50\pm 0.13$~mag. The DEB technique has been applied to a similar
system in M31 
(ref.\ 41)
and in that case, the relative
distance modulus between M31 and the LMC is in excellent agreement
with other methods.  
%

\subsection{Modelling the O-star wind and measuring the photospheric
X-ray eclipse width in M33 X-7:}

The mean X-ray luminosity of  M33 X-7 is 
$8\times 10^{37}$ erg s$^{-1}$.  For an
inclination of $75^{\circ}$, 
and using a limb darkening law\cite{chandra} 
given by
$I=I_0(0.5 + 0.75 \cos i )$, 
the isotropic luminosity is $2.2\times 10^{38}$
erg s$^{-1}$.
For a nominal 10\% accretion efficiency, this corresponds to a mass
accretion rate of $1.7\times 10^{18}$ g s$^{-1}
=2.7\times 10^{-8}\,M_{\odot}$ yr$^{-1}$.
We use this estimate to constrain the O-star wind.

\begin{figure}
\centerline{\includegraphics[scale=0.85,angle=0]{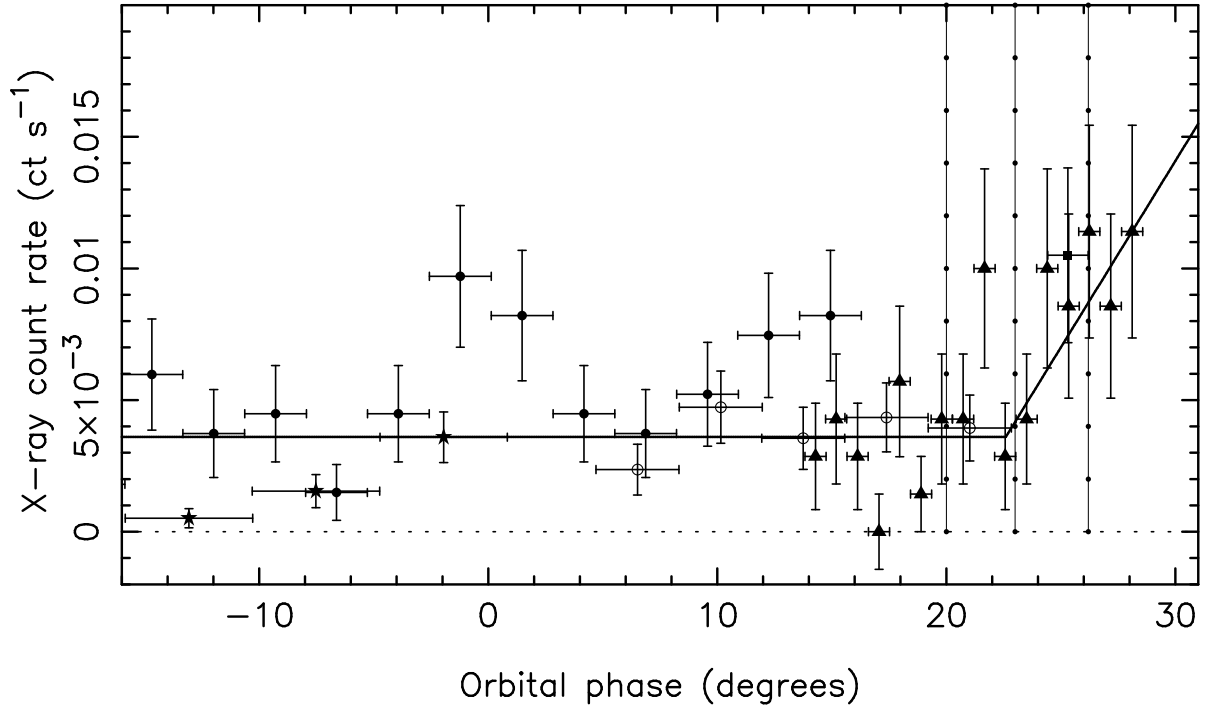}}
\caption{{\bf $\vert$
Determination of the X-ray eclipse angle 
\boldmath{$\Theta$}.}  The portion
of the ACIS light curve from Figure 2 near the egress phase is shown
with phase expressed in units of degrees.  The horizontal bars represent
the phase interval of the integration.
The vertical error bars are  $1\sigma$ (s.d.) statistical.
Different plotting symbols represent different ACIS Observation IDs:
\#6378 filled circles;
\#6382 filled squares;
\#7171 filled triangles;
\#7196 filled stars; and
\#7199 open circles.
The thick solid line is the model
used to determine $\Theta$ and is given by $y(x)=c_1$ for $x\le x_1$ and
$y(x)=(c_2-c_1)/(x_2-x_1)x+c_1$ for $x>x_1$.  The vertical lines
denote, from left to right, eclipse widths of $40^{\circ}$,
$46^{\circ}$, and $53^{\circ}$, respectively.
For the full duration of
the eclipse by the photosphere of the
O-star, we adopt $\Theta = 46^{\circ}$.
\label{Xeclipse}}
\end{figure}

The X-ray intensity following egress increases steeply by roughly a
factor of 15 ($\approx 0.01$ counts s$^{-1}$ versus $\approx 0.15$
counts s$^{-1}$; Supplementary Figure \ref{Xeclipse}).  
Using the out-of-eclipse spectral
models 
in ref.\ 5
and a metalicity of 10\% of the
solar value,
the column density of gas required to attenuate the X-ray
flux by a factor of 15 is $\approx 0.7$ g cm$^{-2}$.

As a function of radius $r$, the velocity $v(r)$ of a radiatively
driven O-star wind has a profile of the form\cite{des05}
\begin{equation}
v(r)=v_{\infty}\left(1-{R_2\over r}\right)^{\beta},
\end{equation}
where $v_{\infty}$ is the asymptotic speed of the wind at large radius,
$R_2$ is the radius of the O-star, and $\beta$ is an index which typically
has a value in the range 0.8 to 1.2.  In the following, we take
$v_{\infty}=2000$ km s$^{-1}$, a typical value for O-stars, $\beta=1$, and
the value of $R_2$ given in Table 2
(main text): $19.6\,R_{\odot}=1.36\times 10^{12}$
cm. With $a=42.4\,R_{\odot}=2.95\times 10^{12}$ cm (Table 2, main text),
the wind velocity at the radius corresponding to the location of
the black hole ($r=a$) is then $v(a)=1081$ km s$^{-1}$.

The mass accretion rate onto the black hole may be roughly estimated
via the Bondi-Hoyle formula 
(e.g.\ ref.\ 44),
\begin{equation}
\dot{M}\approx 4\pi \rho(a)v(a)\left[{2GM\over v^2(a)}\right]^2
 = 1.75\times 10^{32}\rho(a)\,{\rm g~s}^{-1},
\end{equation}
where $\rho(a)$ is the density of the wind at $r=a$.
This formula is only approximately correct and we should include an
unknown coefficient to allow for uncertainties.  However, numerical
simulations\cite{taa88}
show that the coefficient is not very different from unity.

Equating the expression for $\dot{M}$ in equation (2) to the previously
estimated mass accretion rate of $1.7\times 10^{18}$ g s$^{-1}$,
we find $\rho(a)\approx 1.4\times 10^{-14}$ g cm$^{-3}$.
The mass loss rate in the O-star wind is then
\begin{equation}
\dot{M_W}\approx 4\pi a \rho(a)v(a)=2.6\times 10^{-6}\,
M_{\odot}\,{\rm yr}^{-1}.
\end{equation}
For comparison, the mass loss rate derived from line-force 
computations\cite{vin01}
is
$\log{\dot{M}_W}=-6.274\pm
0.226$
($5.32\times 10^{-7}\,M_{\odot}\,{\rm yr}^{-1}$), 
assuming an O-star luminosity and mass given in Table  2
(main text), a temperature between 34,000 and 36,000 K, a metalicity
between 0.1 and 0.3 times the solar value, and 
$2.0 \le v_{\infty}/v_{\rm esc}\le 2.6$.  
This mass loss rate is a factor of 5 smaller
than the value we derived.  However, the plots that compare the
predicted mass loss rates with observed values for LMC and SMC stars
(Figs.\ 6 and 7 in ref.\ 46) have a
scatter of up to 0.6 dex (e.g.\ nearly a factor of 4).  We therefore
do not consider the difference between the mass loss rate we derived
and the one derived from the line-force computations significant.
Finally, using $\dot{M_W}=2.6\times 10^{-6} \, M_{\odot}\,{\rm yr}^{-1}$,
the wind density at radius $r$ is
\begin{equation}
\rho(r)\approx
{\dot{M_W} \over 4\pi r^2 v(r)}={6.6\times 10^{10}\over
r(r-R_2)} \,{\rm g~cm}^{-3}.
\end{equation} 
We have assumed a spherically symmetric wind, which is perhaps
not a very good approximation\cite{blo95}, 
but it is hard to know how to improve on this.

The photosphere of the O-star is located at a radius $R_p$
such that the radial column density from infinity down to $R_p$
is equal to $0.54$ g cm$^{-2}$, which is the column
depth to the $\tau=1$ surface in the best-fitting model atmosphere:
\begin{equation}
\int_{R_p}^{\infty}\rho(r)dr=0.54\,{\rm g~cm}^{-2}.
\end{equation}
Solving, we find that $R_p$ is extremely close to the O-star radius of
$19.6\,R_{\odot}$.  Thus, the presence of the wind has a negligible effect
on the radius of the photosphere.

The X-ray eclipse light curve, however, is strongly affected by the
wind.  Let us define the critical radius of the eclipse
$R_e$ to be the impact parameter at which the line-of-sight
column density is equal to 0.7 g cm$^{-2}$.  Thus
\begin{equation}
\int_{-Z_x}^{\infty}\rho\left(\sqrt{R_e^2+z^2}\right)dz=0.7\,{\rm g~cm}
^{-2},
\end{equation}
where $Z_x=\sqrt{a^2-R_2^2}$ gives the position of the X-ray source.
Solving this condition numerically, we obtain 
an eclipse width of $\Theta=51^{\circ}$.

There are uncertainties in many of the parameters we have used.  Let
us allow for all the uncertainties by varying $\dot{M_W}$ by a factor
of 1.5 either way around the estimate in equation (3).  Over this
entire range of $\dot{M_W}$, $R_p$ continues to remain very close to
$R_2$, so the wind has no effect on the photosphere.  At the lower end
of the range, we obtain for the eclipse radius $\Theta=49^{\circ}$,
while at the upper end, we find $\Theta=55^{\circ}$.  The observed
width of $\Theta=53\pm 2.2^{\circ}$ is bracketed by these values.

According to this model, the photosphere subtends an angle of
$46.3^{\circ}$ in the light curve, while the eclipse, which is defined
by the positions at which the X-ray flux falls by a factor of 15,
subtends an angle in the range $49^{\circ} - 55^{\circ}$.  Thus, the
ratio of the eclipse angle to the photospheric angle lies in the range
1.06 to 1.19.  Turning this around, since the observed eclipse angle
is $53^{\circ}$, we estimate the photospheric angle to lie in the
range $44^{\circ}-50^{\circ}$.

The X-ray light curve itself suggests
an angle close to $46^{\circ}$ (Supplementary 
Figure \ref{Xeclipse}).  
We fitted
two line segments to the ACIS count rate data and found that the
transition from the flat bottom to egress occurs between
$\Theta=40^{\circ}$ ($\pm1^{\circ})$ and $\Theta=46^{\circ}$
($\pm1^{\circ})$, depending on how the data are binned and how they are
fitted (i.e., $\chi^2$ versus median fitting).
Thus, in modelling the
binary system, we set $\Theta = 46\pm1^{\circ}$ (Table 2, main text).
Although we are confident that
$\Theta=46^{\circ}$ is the appropriate value to use, we nevertheless
show below that derived geometrical parameters change only modestly for
values of $\Theta$ between $40^{\circ}$ and $50^{\circ}$.

\subsection{Modelling and error estimation:}

We used the $\chi^2$ statistic to evaluate the goodness-of-fit
between the observed light and velocity curves and their model 
counterparts:
\begin{eqnarray}
\chi^2_{\rm light} &=& \sum_{i=1}^{30}{y(x_i;a_1...a_{10})-y_i^B\over 
     \sigma_i^2} 
 + \sum_{i=1}^{70}{y(x_i;a_1...a_{10})-y_i^V\over 
     \sigma_i^2} \nonumber \\
& & + \sum_{i=1}^{39}{y(x_i;a_1...a_{10})-y_i^{g^{\prime}}\over 
     \sigma_i^2}   +
 \sum_{i=1}^{24}{y(x_i;a_1...a_{10})-y_i^{r^{\prime}}\over 
     \sigma_i^2};  \\
\chi^2_{\rm RV} &= &  \sum_{i=1}^{20}{y(x_i;a_1...a_{10})-y_i^{\rm RV}\over 
     \sigma_i^2}. 
\end{eqnarray}
Here, the notation 
$y_i^B$ 
$y_i^V$ 
$y_i^{g^{\prime}}$ 
$y_i^{r^{\prime}}$ 
are the observed $B$, $V$, $g^{\prime}$, and $r^{\prime}$ magnitudes at
orbital phase $x_i$, respectively,
$\sigma_i$ is the uncertainty on the measurement at 
$x_i$,
and $y(x_i;a_1...a_{10})$ is the model value
at  $x_i$.  We also have three observed quantities that constrain
the available parameter space, namely the duration of the X-ray eclipse
$\Theta$, the radius of the secondary star $R_2$, and the projected
rotational velocity of the secondary star $V_{\rm rot}\sin i$.  
For each model that is computed, these three quantities can be determined.
Hence, there are three additional contributions to
$\chi^2$:
\begin{eqnarray}
\chi^2_{\Theta} &=& {\Theta(a_1...a_{10})-46\over 1^2}, \\
\chi^2_{R_2} &=& {R_2(a_1...a_{10})-19.6\over 0.9^2}, \\
\chi^2_{\rm rot} &=& {V_{\rm rot}\sin i(a_1...a_{10})-250\over 7^2}. 
\end{eqnarray}
Our total $\chi^2$ is then 
\begin{equation}
\chi^2_{\rm total} = \chi^2_{\rm light} + \chi^2_{\rm RV}
+ \chi^2_{\Theta}+ \chi^2_{R_2} + \chi^2_{\rm rot}.
\end{equation}
In any optimisation procedure, there is always the issue of assigning
relative weights to different data sets.  After our extensive initial searches
of parameter space resulted in a good solution, we scaled the uncertainties
on each data set (e.g., four light curves and one radial velocity curve)
so that the total $\chi^2$ of the fit  was equal to $N-1$
for each data set separately.  The required scalings and number of
observations for each data set were
0.8715 for 30 observations in $B$,
0.9505 for 70 observations in $V$,
1.0064 for 39 observations in $g^{\prime}$,
1.0449 for 24 observations in $r^{\prime}$,
and 0.9452 for 22 radial velocity observations.
After the scaling, the optimiser codes were
run again to produce the final set of parameters.  
The resultant $\chi^2$ of the best-fitting
solution is $\approx 180$.

\begin{figure}
\centerline{\includegraphics[scale=0.7,angle=-90]{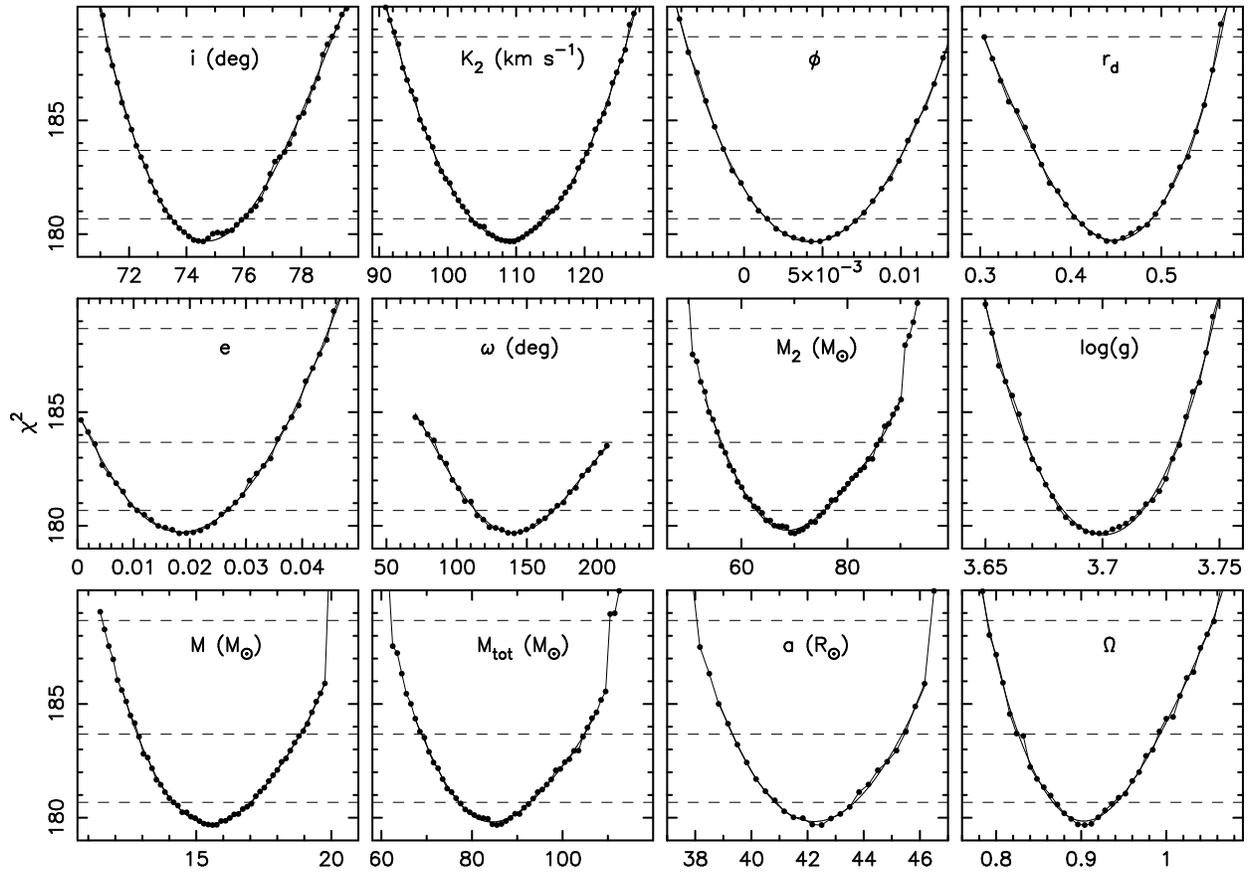}}
\caption{{\bf $\vert$ 
\boldmath{$\chi^2$} vs.\ parameter of interest.}  Projections of
the $\chi^2$ hypersurface projected along various axes are shown
(filled circles).  The $x$-axes give the value of either a fitted parameter
(e.g.\ the inclination) or a derived parameter (e.g.\ the black hole mass),
and the $y$-axes gives the
optimal $\chi^2$ value when the corresponding $x$-axis parameter is fixed at
its particular value.
The dashed lines denote the formal $1\sigma$,
$2\sigma$, and $3\sigma$ confidence intervals.\label{chi}}
\end{figure}

We computed the confidence limits of the fitted and derived
parameters using a brute force method.  
When a model was computed, we saved the value of fitted parameters, the
derived parameters (e.g., the mass of the compact object, surface gravity
of the companion, etc.), and the $\chi^2$ of the fit.  
After a suitably large number
of runs of the genetic optimiser 
and the grid search optimiser, we computed models and $\chi^2$ 
values almost everywhere near the global 
$\chi^2$ minimum in the 10-dimensional parameter space.  It was then
a simple matter to project one dimension of the $\chi^2$ hypersurface
along any parameter of interest.  Supplementary 
Figure \ref{chi}
shows
$\chi^2$ vs.\ parameter value for twelve fitted and derived parameters.
A sixth order polynomial was fitted to each curve to determine the
value of the parameter at the minimum $\chi^2$, and to compute
the formal 1, 2, and $3\sigma$ confidence intervals (taken
to be the range in the parameter needed to make $\chi^2=
\chi^2_{\rm min}+1$, $\chi^2_{\rm min}+4$, and
$\chi^2_{\rm min}+9$, respectively).  Each of the parameters shown
in the figure are well determined, since the curves have smooth
shapes with clear minima.  The non-zero eccentricity is significant
at about the $2\sigma$ level, as is the indication of non-synchronous
rotation (e.g.\ the parameter $\Omega\equiv P_{\rm rot}/P_{\rm orb}$
is different than 1.0 by $\approx 2\sigma$).  On the other hand,
models with no accretion disk are ruled out at the level of several
$\sigma$ since the best-fitting model with no accretion disk
has a $\chi^2$ of more than 210.


As a check, extensive fits were made assuming different values
of the X-ray eclipse width $\Theta$.  Supplementary Table 1
shows the best-fitting values of the inclination, secondary star mass,
black hole mass, and $\chi^2$ for various assumed values of
$\Theta$.   Over the whole range of $\Theta$ ($40^{\circ}-50^{\circ}$), the 
inclination changes from
$71.5^{\circ}$ to $77.3^{\circ}$, 
and the black hole mass only changes from
$15.97\,M_{\odot}$ to $15.35\,M_{\odot}$.

\subsection{Comparison of ellipsoidal model with additional
light curves:}

\begin{figure}
\centerline{\includegraphics[scale=0.63]{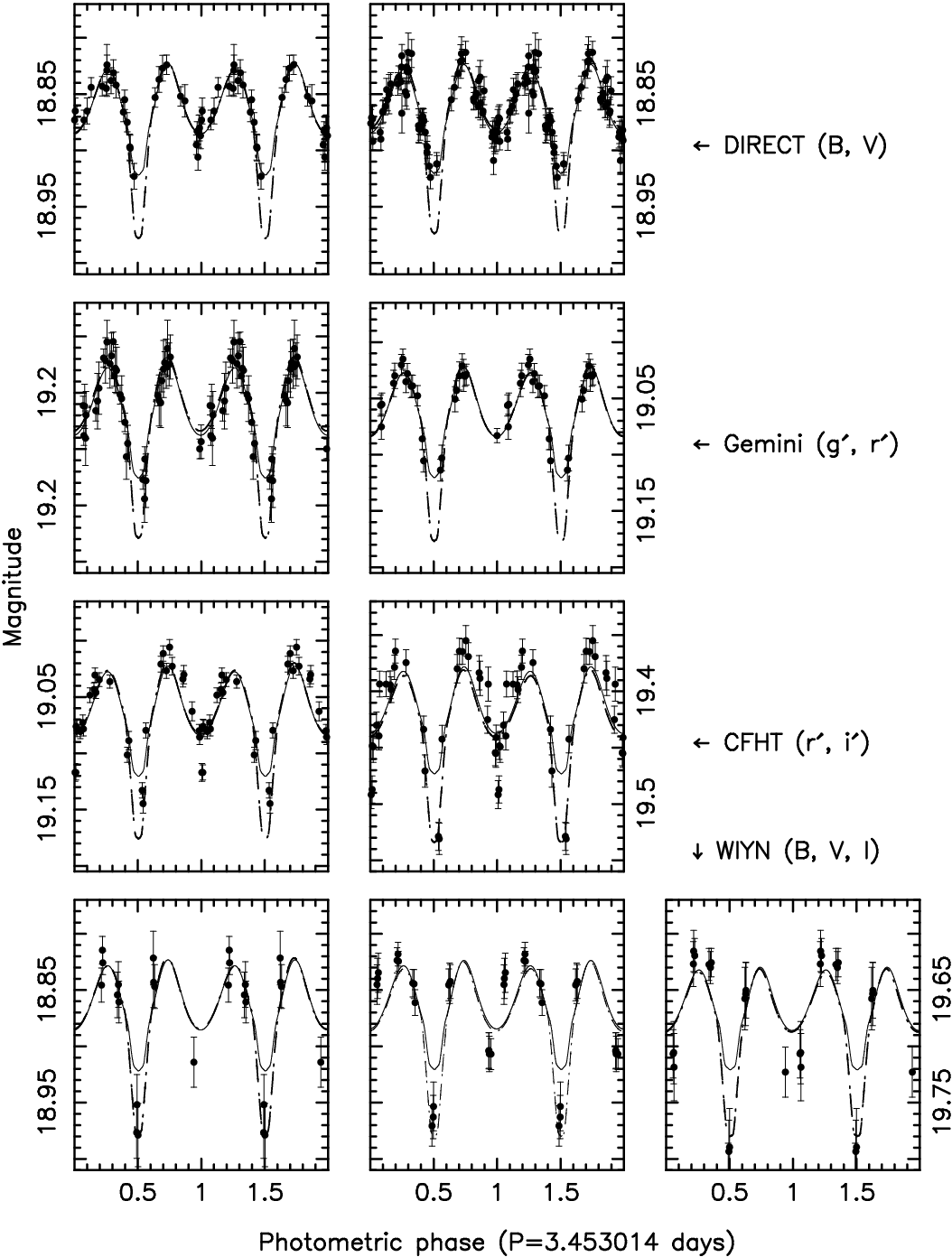}}
\caption{{\bf $\vert$
Phased light curves and ellipsoidal models.}
All of the available optical light curves phased on the
orbital period are shown with the ellipsoidal model fitted to the
DIRECT $B$ and $V$ and the Gemini $g^{\prime}$
and $r^{\prime}$ light curves (solid lines).
The dashed line shows an ellipsoidal model with a larger accretion disk
that qualitatively explains the deeper minima near phase 0.5 in
the CFHT $r^{\prime}$ and $i^{\prime}$ and the WIYN
$B$, $V$, and $I$ light curves compared to the DIRECT and Gemini 
counterparts.  The error bars are $1\sigma$ (s.d.) statistical.
\label{lcfig2}}
\end{figure}

Supplementary Figure 
\ref{lcfig2}
shows all of the available optical light
curves of M33 X-7 phased on the X-ray ephemeris$^5$: 
$B$ and $V$ light curves from the DIRECT survey
(see ref.\ 10),
$g^{\prime}$ and 
$r^{\prime}$ from the Gemini North Telescope,
$r^{\prime}$ and $i^{\prime}$ from the Canada-France-Hawaii Telescope
(CFHT) M33 Variability Survey 
(see ref.\ 11)
and
$B$, $V$, and $I$ from WIYN.  The CFHT and WIYN light curves have
anomolously faint points near phase 0 (1.e., near X-ray eclipse)
that we cannot explain (there are no obvious defects in the images
in question).  These same
light curves also have deeper minima near phase 0.5
(when the X-ray source is in front of the O-star companion),
which, unlike the low points near phase 0, can easily be explained 
qualitatively by increasing the radius of the accretion disk
from 45\% of the black hole's Roche lobe radius to
$\approx 80\%$.  The out-of-eclipse X-ray flux of M33 X-7 is
variable, so an accretion disk with a changing radius is perhaps not
unexpected.  Alternatively, if the plane of the accretion disk is not
exactly parallel to the plane of the binary orbit, and the disk
precesses, then its cross sectional area on the plane of the sky
would change.  This would result in changes in the depth of the light
curve minimum near phase 0.5.

We note that in the DIRECT, Gemini, and CFHT light curves
the maximum near phase 0.75 is consistently higher than the
maximum near phase 0.25.  The small orbital eccentricity in our model
explains the slight difference between the heights of the maxima.

\section{Supplementary notes}

\spacing{1}

\newpage

\section{Supplementary figures, legends, and table}

\newpage

\newpage

\newpage

\newpage

\begin{table}
\begin{tabular}{rrrrr}
\hline
$\Theta$ (deg)  & $i$ (deg) & $M_2$ ($M_{\odot}$) & $M$ ($M_{\odot}$) &
         $\chi^2 $ \\
\hline
\hline
40.0 & 71.50 & 69.72 &  15.97 & 181.850 \\
41.0 & 71.85 & 69.70 &  15.90 & 181.132 \\
42.0 & 72.36 & 69.79 &  15.79 & 181.018 \\
43.0 & 73.05 & 69.89 &  15.82 & 180.460 \\
44.0 & 73.51 & 70.15 &  15.77 & 180.365 \\
45.0 & 74.05 & 70.08 &  15.73 & 179.914 \\
\bf 46.0 & \bf 74.62 & \bf 70.00 &  \bf 15.65 & \bf 179.675 \\
47.0 & 75.51 & 70.24 &  15.59 & 179.637 \\
48.0 & 75.84 & 70.28 &  15.49 & 179.542 \\
49.0 & 77.05 & 70.13 &  15.39 & 179.546 \\
50.0 & 77.27 & 70.08 &  15.35 & 179.938 \\
\hline
\end{tabular}
\caption{Results for different values of X-ray eclipse width $\Theta$.  
The best-fitting values of the inclination $i$, mass of the secondary
star $M_2$, the black hole mass $M$, and the $\chi^2$
of the fit
for various assumed values of the X-ray eclipse width $\Theta$
are shown.
The row in boldface is the one adopted in the main text.
}
\end{table}
  

\begin{thebibliography}{1}

\bibitem[26]{kim02}
Kim, M., Kim. E., Lee, M. G., Sarajedini, A. \& Geisler, D.
  Determination of the distance
  to M33 based on the tip of the red giant branch and the red clump.
  {\em Astron.\ J.} {\bf 123}, 244-254 (2002).

\bibitem[27]{sak00}
   Sakai, S., Zaritsky, D., \& Kennicutt, J. C.
   The tip of the red giant branch distance to the Large Magellanic Cloud.
   {\em Astron.\ J.} {\bf 119}, 1197-1204 (2000).


\bibitem[28]{mac01}
Macri, L. Ph.D.\ Thesis, Harvard University (2001).

\bibitem[29]{uda99}
   Udalski, A., Szymanski, M., Kubiak, M., Pietrzynski, G.,
   Soszynski, I., Wozniak, P. \&  Zebrun, K.
   The Optical Gravitational Lensing Experiment.
   Cepheids in the Magellanic Clouds. III. period-luminosity-color
   and period-luminosity relations of classical Cepheids.
   {\em Ac.\ A.} {\bf 49}, 201-221 (1999).

\bibitem[30]{mac06}
   Macri, L. M.,  Stanek, K. Z., Bersier, D.,
   Greenhill, L. J. \&  Reid, M. J.
   A new Cepheid distance to the maser-host galaxy NGC 4258
   and its implications for the Hubble constant.
   {\em Astrophys.\ J.} {\bf 652}, 1133-1149 (2006).

\bibitem[31]{sak04}
  Sakai, S.,  Ferrarese, L.,  Kennicutt, R. C. \&  Saha, A.
  The effect of metallicity on Cepheid-based distances
  {\em Astrophys. J.} {\bf 608}, 42-61 (2004).

\bibitem[32]{sar06}
  Sarajedini, A.,
  Barker, M. K.,  Geisler, D.,  Harding, P. \&  Schommer, R.
  RR Lyrae variables in M33. I. Evidence for a field halo population.
  {\em Astron.\ J.} {\bf 132} 1361-1371 (2006).

\bibitem[33]{cle03}
  Clementini, G.,  Gratton, R.,  Bragaglia, A.,
  Carretta, E.,  Di Fabrizio, L. \&  Maio, M.
  Distance to the Large Magellanic Cloud: The RR Lyrae stars.
  {\em Astron.\ J.} {\bf 125} 1309-1329 (2003).

\bibitem[34]{cha99}
  Chaboyer, B., Green, E. M. \& Liebert, J.
  The age, extinction, and distance of the old, metal-rich
  open cluster NGC 6791.
  {\em Astron.\ J.} {\bf 117}, 1360-1374.

\bibitem[35]{gra03}
   Gratton, R. G., Bragaglia, A., Carretta, E.,
   Clementini, G.,  Desidera, S.,  Grundahl, F. \& Lucatello, S.
   Distances and ages of NGC 6397, NGC 6752 and 47 Tuc.
   {\em Astron.\ Astrophys.} {\bf 408}, 529-543 (2003).

\bibitem[36]{uda00}
   Udalski, A.
   The Optical Gravitational Lensing Experiment: red clump stars
   as a distance indicator.
   {\em Astrophys.\ J.} {\bf 531}, L25-L28 (2000).

\bibitem[37]{pop00}
   Popowski, P. Clump giant distance to the Magellanic Clouds and
   anomalous colors in the galactic bulge.
   {\em Astrophys.\ J.} {\bf 528}, L9-L12 (2000).

\bibitem[38]{fit03}
  Fitzpatrick, E. L.,  Ribas, I.,  Guinan, E. F.,
  Maloney, F. P. \& Claret, A.
  Fundamental properties and distances of Large Magellanic Cloud eclipsing
  binaries. IV. HV 5936.
  {\em Astrophys. J.} {\bf 587}, 685-700 (2003).

\bibitem[39]{her99}
 Herrnstein, J. R., Moran, J. M., Greenhill, L. J.,
 Diamond, P. J., Inoue, M.,  Nakai, N.,
 Miyoshi, M.,  Henkel, C. \& Riess, A.
 A geometric distance to the galaxy NGC4258 from
 orbital motions in a nuclear gas disk. {\em Nature} {\bf 400},
 539-541 (1999).

\bibitem[40]{bon06}
  Bonanos, A. Z. {\em et al.} The first
  DIRECT distance determination to a detached eclipsing binary in
  M33.
  {\em Astrophys.\ J.} {\bf 652}, 313-322 (2006).

\bibitem[41]{rib05}
  Ribas, I. {\em et al.} First determination
  of the distance and fundamental properties of an
  eclipsing binary in the Andromeda Galaxy.
  {\em Astrophys.\ J.} {\bf 635}, L37-L40 (2005).

\bibitem[42]{chandra}
Chandrasekhar, S. {\em Radiative Transfer}, (Clarendon Press,
Oxford, 1950).

\bibitem[43]{des05}
Dessart, L. \& Owocki, S. Inferring hot-star-wind acceleration from
line profile variability. {\em Astron.\ Astrophys.} {\bf 432}, 281-294
(2005).

\bibitem[44]{sha83}
Shapiro, S. L., Teukolsky, S. A. {\em Black holes, white dwarfs,
and neutron stars: The physics of compact objects}, (Wiley-Interscience,
New York, 1983).

\bibitem[45]{taa88}
Taam, R. E. \& Fryxell, B. A. On nonsteady accretion in stellar
wind-fed X-ray sources. {\em Astrophys.\ J.}, {\bf 327}, L73-L76
(1988).

\bibitem[46]{vin01}
Vink, J. S., de Koter, A., \& Lamers, H. J. G. L. M. Mass-loss
predictions for O and B stars as a function of metallicity.
{\em Astron.\ Astrophys.} {\bf 369}, 574-588 (2001).

\bibitem[47]{blo95}
Blondin, J. M. \& Woo, J. W.  Wind dynamics in SMC X-1. 1:
Hydrodynamic simulation. {\em Astrophys.\ J.}, {\bf 445}, 889-895
(1995).



\end{thebibliography}
\end{document}